# A Movable Valley Switch Driven by Berry Phase in Bilayer Graphene Resonators

Yi-Wen Liu[1,§], Zhe Hou[2,§], Si-Yu Li[1,§], Qing-Feng Sun[2,3,4,*], and Lin He[1,5,*]

[1] Center for Advanced Quantum Studies, Department of Physics, Beijing Normal University, Beijing, 100875, People's Republic of China

[2] International Center for Quantum Materials, School of Physics, Peking University, Beijing, 100871, China

[3] Collaborative Innovation Center of Quantum Matter, Beijing 100871, China

[4] Beijing Academy of Quantum Information Sciences, West Bld. #3, No. 10 Xibeiwang East Road, Haidian District, Beijing 100193, China

[5] State Key Laboratory of Functional Materials for Informatics, Shanghai Institute of Macrosystem and Information Technology, Chinese Academy of Sciences, 865 Changning Road, Shanghai 200050, People's Republic of China

[§]These authors contributed equally to this work.
*Correspondence and requests for materials should be addressed to Qing-Feng Sun (e-mail: sunqf@pku.edu.cn) and Lin He (e-mail: helin@bnu.edu.cn).

**Since its discovery, Berry phase has been demonstrated to play an important role in many quantum systems. In gapped Bernal bilayer graphene, the Berry phase can be continuously tuned from zero to $2\pi$, which offers a unique opportunity to explore the tunable Berry phase on the physical phenomena. Here, we report experimental observation of Berry phases-induced valley splitting and crossing in moveable bilayer graphene p-n junction resonators. In our experiment, the bilayer graphene resonators are generated by combining the electric field of scanning tunneling microscope tip with the gap of bilayer graphene. A perpendicular magnetic field changes the Berry phase of the confined bound states in the resonators from zero to $2\pi$ continuously and leads to the Berry phase difference for the two inequivalent valleys in the bilayer graphene. As a consequence, we observe giant valley splitting and unusual valley crossing of the lowest bound states. Our results indicate that the bilayer graphene resonators can be used to manipulate the valley degree of freedom in valleytronics.**

Berry phase, the geometric phase accumulated over a closed loop in parameter space during an adiabatic cyclic evolution [1,2], has attracted much attention because its important role in determining physical properties of many quantum systems, such as graphene systems [3-9]. In graphene monolayer, the Berry phase can take two quantized values, 0 or π, and usually it only takes either 0 or π because changing the Berry phase of quasiparticles is experimentally challenging. Until very recently, it was demonstrated explicitly that the Berry phase of quasiparticles confined in circular graphene monolayer resonators can be tuned from 0 to π by magnetic fields. The jump of the Berry phases between the two quantized values leads to a sudden and large jump in energy of the quasi-bound states [10]. Such a result reveals the close relationship between the Berry phase and the electronic properties of graphene systems [9-11]. In graphene systems, gapped graphene bilayer is quite unique because that the Berry phase of its quasiparticles can be continuously tuned from zero to 2π [12-24]. Therefore, it provides an unprecedented platform to explore the effects of continuous tunable Berry phase on the physical phenomena.

In this Letter, we experimentally study confined bound states in moveable bilayer graphene p-n junction resonators and observe Berry phases-induced valley splitting and crossing of the bound states. By combining the electric field of scanning tunneling microscope (STM) tip with the gap of bilayer graphene, we realize the bilayer graphene p-n junction resonator, which enables us to detect valley splitting at single-electron level [25-28]. Our experiment indicates that a perpendicular magnetic field generates giant valley splitting and unusual valley crossing of the lowest bound states, attributing to the continuous tunable Berry phase of the bound states in the resonators.

Our experiments were carried out on Bernal bilayer graphene by using a high-magnetic-field STM at $T$ = 4.2 K (see Supplemental Material for details [29]). The Bernal bilayer graphene was grown on Cu foils by low pressure chemical vapor deposition (LPCVD) method and was transferred from the Cu foils onto single-crystal SrTiO$_3$ substrates for further STM characterizations [30-32]. Figure 1(a) shows a representative STM image of the bilayer graphene and the triangular contrast in the atomic-resolved STM image (inset of Fig. 1(a)) arises from the A/B atoms' asymmetry in the Bernal bilayer graphene, as observed previously [33,34]. To further identify the stacking order of the adjacent bilayer graphene, scanning tunneling spectroscope (STS) measurements in various magnetic fields were carried out, as shown in Fig. 1(b). In zero magnetic fields, the tunneling spectrum exhibits a finite gap ~ 46 meV, which is generated by inversion symmetry breaking of the adjacent two layers induced by the substrate. In high magnetic fields, the spectra show well-defined Landau quantization of massive Dirac fermions (Fig. 1(c) and see Fig. S1 for details of analysis [29]), as observed in Bernal bilayer graphene [33,34]. Figure 1(d) shows a typical STS map, which can reflect the local density of states (LDOS) at the atomic level, recorded at energy of the conduction-band edge of the bilayer graphene. Obviously, it also exhibits the triangular contrast due to the stacking order and inversion symmetry breaking of the Bernal bilayer graphene. All the above measurements demonstrate explicitly that the studied graphene bilayer are Bernal stacked.

Besides the well-defined Landau levels of massive Dirac fermions and the low-energy gap, we also observe quadruplet of charging peaks in the tunneling spectra, as shown in Fig. 1(b). The emergence of quadruplet of charging peaks indicates the formation of the edge-free graphene quantum dots (p-n junction resonators) beneath the STM tip. The edge-free graphene quantum dots (QDs) are generated by combining the electric field of the STM tip with a finite gap [25-28], as schematically shown in Fig. 2(a). The probing STM tip acts as a top gate and generates band bending below the tip. Combining with the bandgap of the Bernal bilayer graphene, the charge carriers are restricted in the QDs to generate confined orbital states. There are four-fold degeneracies (two for spin and two for valley) for electronic states in gapped bilayer graphene. Therefore, every single orbital state of the edge-free QD could be occupied by four electrons (See Fig. 2(b)). Due to the small capacitance of the QDs, the charging energy $E_c$ is large so that we can observe a series of quadruplets of charging peaks in the STS spectra (See Fig. 1(b), here $E_c = e^2/C$ with $C$ the capacitances of the edge-free QD). We define the discrete single-electron levels as $E_1$ to $E_4$ and the energy difference between adjacent peaks is $E_c$, as shown in Fig. 2(b). When the valley degeneracy is lifted (the valley splitting is $E_v$), every single quadruplet of the confined orbital state will be divided into two doublets and the energy difference between $E_2$ and $E_3$ will become $E_c + E_v$, as schematically shown in Fig. 2(b). Therefore, by using the edge-free QDs, we can detect valley splitting of the Bernal graphene bilayer at single-electron level [25-28].

Our experimental result shown in Fig. 1(b) indicates that there is unusual valley-related phenomenon in the bilayer graphene QD as a function of magnetic fields. To clearly show this, we plot the four charging peaks in the spectra as a function of magnetic fields in Fig. 2(c). The energy spacing of the four charging peaks $\Delta E$ in the QD can be directly deduced from the voltage difference $\Delta V_{tip}$ acquired from the charging peaks by using $\Delta E = \eta e \Delta V_{tip}$ with $\eta$ as the tip lever arm, which can be roughly estimated according to the position of the conduction-band edge and the onset of charging peaks in the STS spectra. The deduced values of the $\Delta E_{12}$, $\Delta E_{23}$ and $\Delta E_{34}$ in different magnetic fields are shown in Fig. 2(d). In zero magnetic field, the $\Delta E_{23}$ is slightly larger than the average value of the $\Delta E_{12}$ and the $\Delta E_{34}$, indicating a small valley splitting $E_v \sim 0.87$ meV in the Bernal bilayer graphene. This zero-magnetic-field valley splitting was also observed in transport measurement through a bilayer graphene QD very recently and was attributed to small potential difference between the two graphene layers [35]. For the magnetic field smaller than 2 T, the $\Delta E_{23}$ increases dramatically with increasing magnetic fields. However, it decreases very quickly with increasing magnetic fields for $B > 2$ T. Similar phenomenon is also observed in the other Bernal bilayer graphene on different substrate (see Fig. S2 [29]). The large energy difference between the $\Delta E_{23}$ and the $\Delta E_{12}$ (or the $\Delta E_{34}$) for $B < 2$ T is attributed to the giant valley splitting in bilayer graphene in the presence of magnetic field. With assuming a Zeeman-like dependence $E = g_v \mu_B B$ ($\mu_B$ is the Bohr magneton), an effective gyromagnetic ratio $g_v$-factor is estimated as about 40 for $B < 2$ T, which is consistent well with that measured in bilayer graphene QDs $\sim 36$ through transport measurement [36]. Similar large $g_v$ for the valley splitting is also observed in monolayer graphene and ABC-stacked trilayer graphene: $g_v \sim 28$ in monolayer graphene [37] and $g_v$-factor ~ 23 in ABC-stacked trilayer graphene [38].

In order to extract the valley splitting in the bilayer graphene QD induced by the magnetic field, we introduce the bare levels $\tilde{E}_i(B) = E_i(B) - E_i(0)$ (with $i$ = 1, 2, 3, 4) to eliminate the effects of the charging energy $E_c$ and zero-magnetic-field valley splitting $E_v$. The levels $\tilde{E}_i$ are quadruplet at $B = 0$ and split into two doublets at the non-zero $B$. The intra-doublet splitting is small and increases linearly with the magnetic field, which is attributed to the spin Zeeman splitting $g_s\mu_B B$ with the $g_s$-factor estimated about 6 (see Fig. S3 [29]). The $g_s$-factor of the spin Zeeman splitting agrees with that obtained very recently in spin splitting of zero LL in graphene monolayer [37]. By eliminating the spin Zeeman splitting, two valley levels are obtained as $\varepsilon_1 = (\tilde{E}_1 + \tilde{E}_2)/2$ and $\varepsilon_2 = (\tilde{E}_3 + \tilde{E}_4)/2$, as shown in Fig. 3. According to Fig. 3, the unusual valley splitting induced by the magnetic field $B$ is clearly observed. For $B < 2$ T, the valley splitting increases quickly as $B$ increases. At about $B = 2$ T, the level $\varepsilon_2$ reaches the maximum. Then for $B > 2$ T, the level $\varepsilon_2$ decreases dramatically and the splitting between $\varepsilon_1$ and $\varepsilon_2$ is almost unchanged with further increasing $B$.

To fully understand the observed unusual valley-related phenomenon in the bilayer graphene QD shown in Fig. 2(c) and Fig. 3, we adopt a bilayer graphene QD model, as recently reported in Ref. [20]. Our analysis indicates that, in the presence of magnetic field, the energy spectra of the bound states in the QD are mainly determined by the Berry phase and the orbital magnetic moment. Consider a circular bilayer graphene QD, the bound eigen-states can be labelled by $(\xi, n, m)$ with $\xi$ the valley index, $n = 0, 1, 2, ...$ the radial quantum number and $m$ the azimuthal (angular momentum) quantum number. At zero magnetic field, the lowest bound levels are $(K, 0, 1)$ and $(K', 0, -1)$, which are mutually time-reversal states. In the semiclassical picture, the motion of particles inside the QD can be decomposed into a rotation in the axial direction with trajectory $C_\theta$ and an oscillation in the radial direction with trajectory $C_r$ on a torus (see Figs. 4(a) and 4(b)). Applying the Einstein-Brillouin-Keller (EBK) formula in the $C_r$ loop [20], we obtain

$$\frac{1}{\hbar}\oint_{C_r} \Pi_r dr = 2\pi(n+\gamma) - \Gamma(C_r) , \qquad (1)$$

which can help us to determine the bound energy inside the QD. Here $\Pi_r$ is the radial momentum, $\gamma$ is the Maslov index and $\Gamma(C_r)$ is the Berry phase the particle accumulates along the loop $C_r$. In small magnetic fields, the radial momentum $\Pi_r$ can be expanded as $\Pi_r = \Pi_r(B=0) + \frac{\partial \Pi_r}{\partial B}|_{B=0} + o(B^2) \approx \Pi_r|_{B=0} + \frac{1}{\Pi_r|_{B=0}}\frac{e\hbar mB}{2}$. Then the EBK quantization formula can be rewritten as:

$$\frac{1}{\hbar}\oint_{C_r}\Pi_r(E,B=0)dr \approx 2\pi(n+\gamma)-\Gamma(C_r)+\frac{1}{\hbar}\int_0^T \mu_M \cdot B dt, \qquad (2)$$

where $T$ is the period the particle spends on travelling around the loop $C_r$, and $\mu_M = \frac{-m\hbar e}{2M_c}$ is the orbital magnetic moment with the effective mass $M_c$. According to Eq. (2), the magnetic field has two effects on the energy spectra of the bound states: (1) inducing a level splitting between $\pm m$ states arising from the orbital magnetic moment in the third term and (2) increasing (decreasing) the Berry phase $\Gamma(C_r)$ for the $m=1$ ($m=-1$) state as shown in Fig. 4(c). The effect of magnetic field on the Berry phase can be explained in Fig. 4(d,e) where the momentum trajectories at different magnetic fields are plotted. The Berry curvature has opposite signs in the two inequivalent valleys. At zero $B$, the momentum trajectories for the two time-reversal related states $(K,n,m)$ and $(K',n,-m)$ enclose the same area but their motion direction are opposite, leading to the Berry phases accumulated for the $(K,n,m)$ and $(K',n,-m)$ states being equal (see Fig.4(b)) [39]. However, for nonzero $B$, the magnetic field can provide an effective Lorentz force that pulls the momentum trajectories in the opposite directions for the two time-reversal related states as shown in Fig. 4(b,e,f), which generates a Berry phase difference for the $(K,n,m)$ and $(K',n,-m)$ states. Since the first term in Eq. (2) has positive correlation with the energy $E$ and does not depend on the valley index and the azimuthal quantum number $m$, therefore, the Berry phase and orbital magnetic moment both have positive [negative] contribution to the energy of $(K',n,-m)$ [$(K,n,m)$] state. As a consequence, we can obtain a large valley splitting with small magnetic fields.

To qualitatively show how the Berry phase and the orbital magnetic moment affect the energy spectra of the bound states in the bilayer graphene QD, we carry out calculations based on the low-energy effective Hamiltonian: $\widehat{H}_\xi = \widehat{H}_{0\xi} + \Delta\tau_z\sigma_0/2 + \begin{pmatrix} U_1(r)\sigma_0 & 0 \\ 0 & U_2(r)\sigma_0 \end{pmatrix}$, where $\widehat{H}_{0\xi}$ is the Hamiltonian of a pristine BLG with $\xi = \pm 1$ denoting the $K(K')$ valley, $\Delta$ is the interlayer bias or the energy gap, $U_{1(2)}(r)$ is layer-dependent electrostatic potentials in the top (bottom) layer, and $r$ is the off-center displacement (see Supplementary Materials for details [29]). The potential is taken into a Gaussian-function form with $U_{1(2)}(r) = U_0 - S_{1(2)}e^{-r^2/\lambda^2}$ ($\lambda$ denotes the radius of the QD). By solving the Schrödinger equation: $\widehat{H}_\xi \Psi_{\xi n,m} = E_{\xi n,m}\Psi_{\xi n,m}$, the bound energy levels $E_{\xi n,m}$ inside the QD is obtained. By taking into account the measured energy gap $\Delta$, the potential difference $\Delta S = S_1 - S_2 = 28$ meV and assuming the QD radius $\lambda = 130$ nm, which is a typical size of an STM generated graphene QD [25-28], we calculate the lowest three bound levels $(K, 0, 1)$, $(K', 0, -1)$ and $(K, 1, 1)$ as a function of magnetic fields, as shown in Fig. 3. Here each level has two spin degeneracies. Obviously, there is a quite large valley splitting between the $(K, 0, 1)$ and the $(K', 0, -1)$. At a critical magnetic field $B_c \sim 2$ T, a crossing between the $(K', 0, -1)$ and the $(K, 1, 1)$ is obtained. In particular, one can see that the theoretical bound levels are in perfect agreement with that obtained in the experiment. Therefore, we can conclude that the large valley splitting is induced by the Berry phase and orbital magnetic moment and the level $\varepsilon_2$ decreasing with increasing magnetic field for $B > 2$ T arises from the crossing of the $(K', 0, -1)$ and the $(K, 1, 1)$ levels. Note that the valley index is sharply changed by slightly tuning the magnetic field near the critical magnetic field $B_c$, which suggests that a $B$-controlled sensitive valley switch can be realized in our QD system. At last, let us consider the spin Zeeman splitting $g_s\mu_B B$ with $g_s = 6$, charging energy $E_c$ and zero-magnetic-field valley splitting $E_v$, and then the lowest four bound states versus the magnetic field are obtained [see the solid curves in Fig. 2(c)]. These theoretical bound states are well consistent with the experiment results, and it clearly indicates that the bend of the states $E_3$ and $E_4$ at $B \approx 2$ T originates from the crossing of the two valley states.

In conclusion, we study confined bound states in moveable bilayer graphene QDs and observe valley splitting and crossing of the bound states at the single-electron level. Our analysis indicates that these interesting valley-related phenomena are driven by the Berry phase and orbital magnetic moment effect in bilayer graphene QDs in the presence of magnetic fields. The result suggests that gapped bilayer graphene is an ideal platform to explore the tunable Berry phase on the physical phenomena.

**Acknowledgments**

This work was supported by the National Natural Science Foundation of China (Grant Nos. 11974050, 11674029, 11921005) and National Key R and D Program of China (Grant No. 2017YFA0303301). L.H. also acknowledges support from the National Program for Support of Top-notch Young Professionals, support from "the Fundamental Research Funds for the Central Universities", and support from "Chang Jiang Scholars Program".

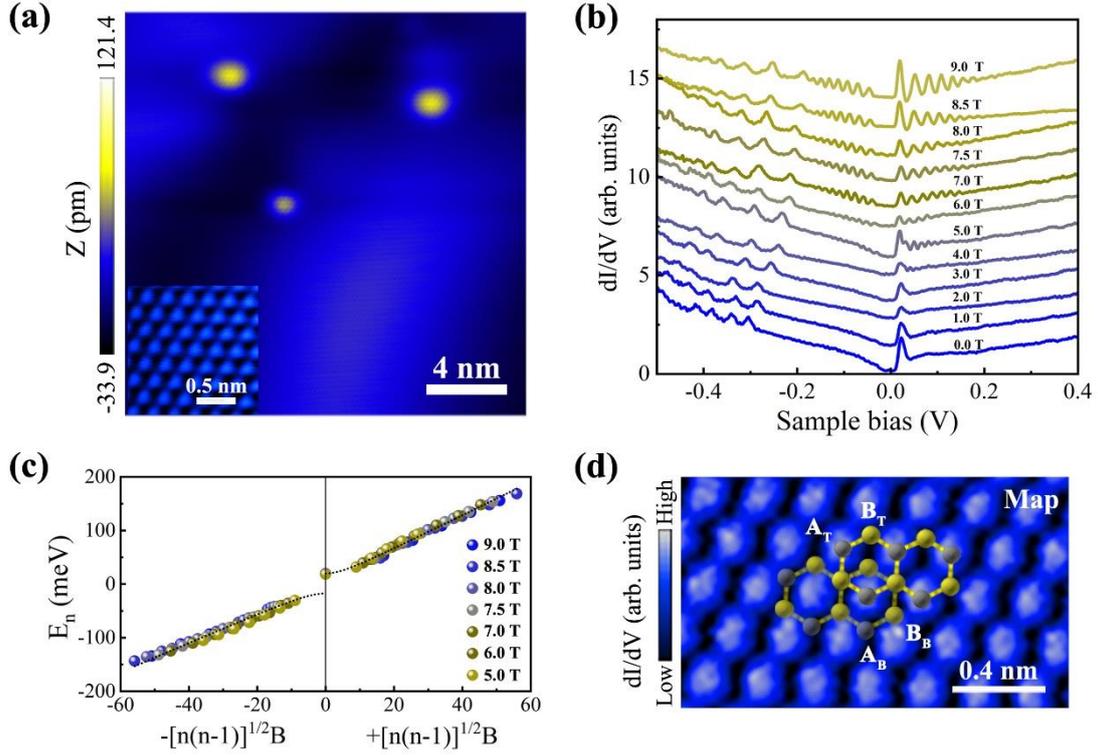

FIG. 1. (a) An STM topographic image of a bilayer graphene region on a STO substrate. The bright dots correspond to defects of the substrate. Inset: atomic resolution STM image of the bilayer graphene showing the triangular contrasting. (b) STS spectra of the Bernal bilayer graphene recorded in different magnetic fields. Around the charge neutrality point, we observed well-defined Landau levels of massive Dirac fermions. The four peaks at high energy correspond to the quadruplets of the confined bound states in the QD beneath the STM tip. (c) The Landau level energies, extracted from panel (b), as a function of $\pm[n(n-1)]^{1/2}B$. The dashed lines are fits to the Landau quantization of massive Dirac fermions in Bernal bilayer graphene. (d) The conductance map recorded at a sample bias of 23 mV. Scale bar: 0.4 nm. The schematic structure of the Bernal bilayer graphene is overlaid on the STS map.

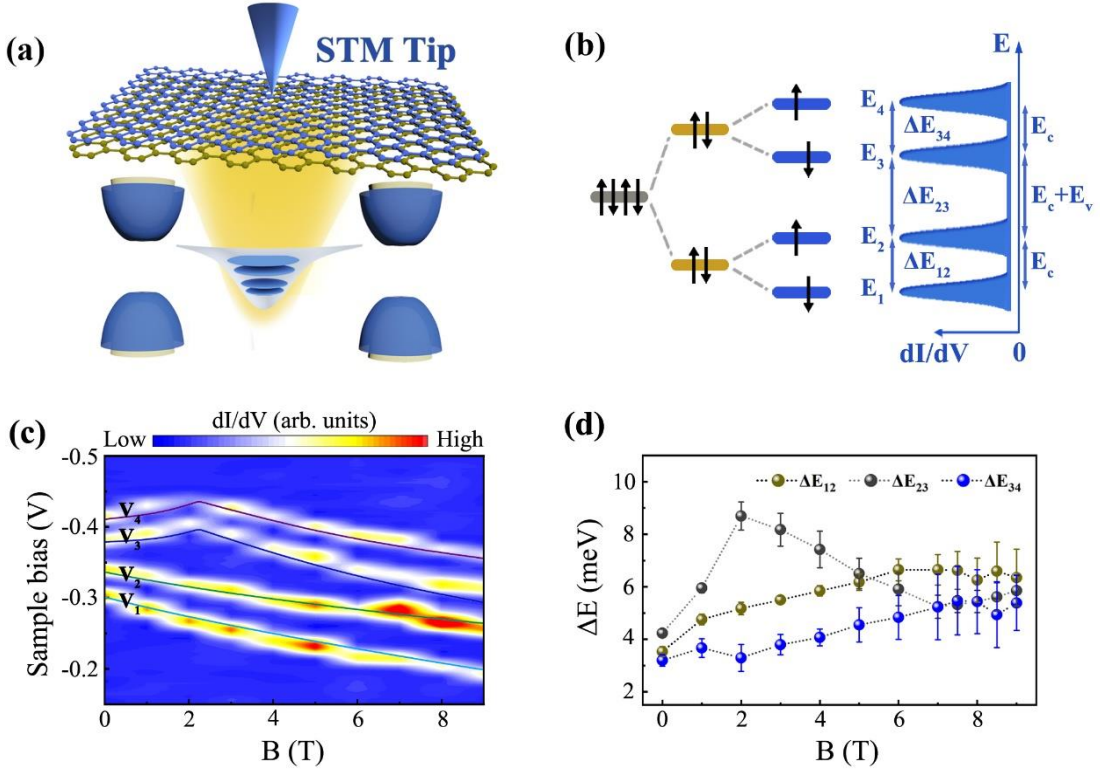

FIG. 2. (a) Schematic of the tip-induced bound states in a bilayer graphene QD. The electrostatic potential of the bilayer graphene beneath the STM tip shifts a little with respect to the surrounding region. (b) A diagram of the magnetic-field-dependent energy splitting for the quadruplets with spin and valley degeneracy in one orbital. A series of charging peaks detected in the spectroscopy are marked with $E_1$ to $E_4$. The electrostatic repulsion among these charging peaks is $E_c$. When the additional valley splitting is involved, the energy spacing between the second and the third peaks becomes $E_c + E_v$. (c) The quadruplets of the confined bound states in the QD measured in different magnetic fields. The sample bias of each bound state is marked with $V_1$ to $V_4$. The solid curves are the theoretical results. (d) The energy spacing between the bound states as a function of the magnetic field in the bilayer graphene QD.

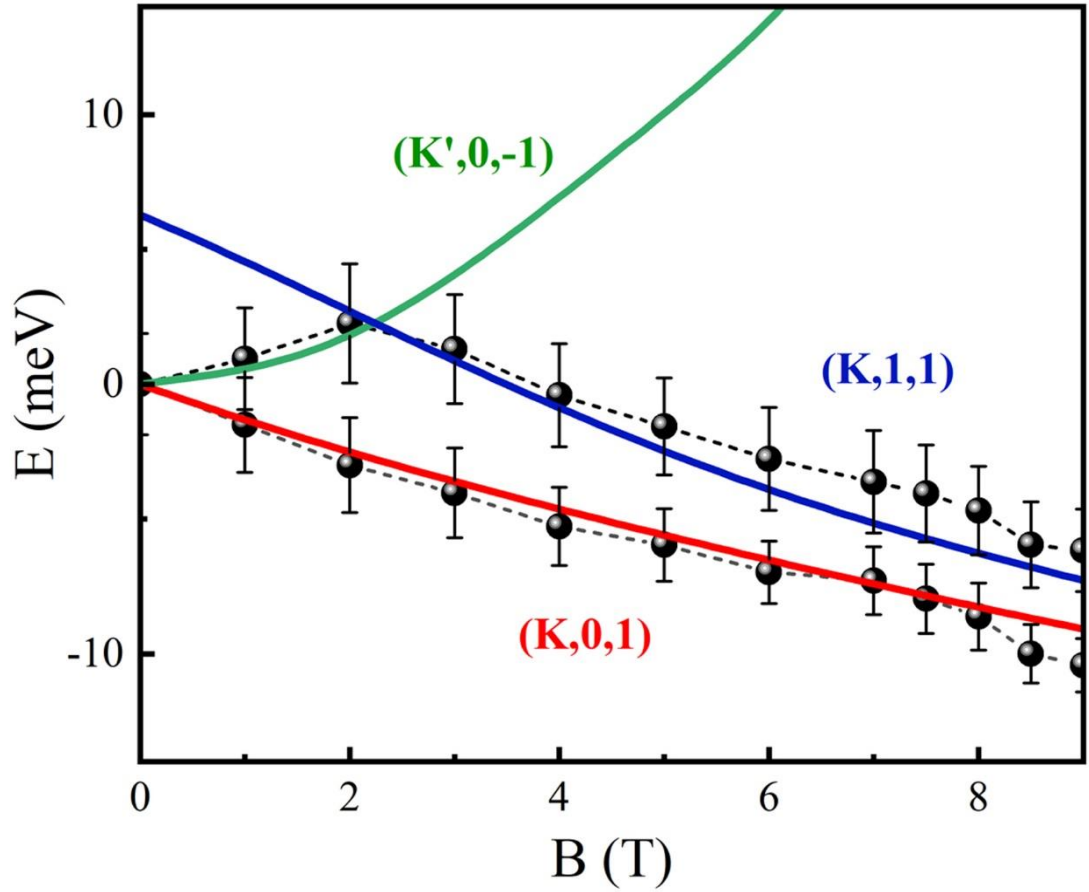

FIG. 3. Comparison of the experimental bound valley levels $\varepsilon_1$ and $\varepsilon_2$ (solid balls) with the theoretically calculated valley levels $(K, 0, 1)$, $(K', 0, -1)$ and $(K, 1, 1)$ (solid lines).

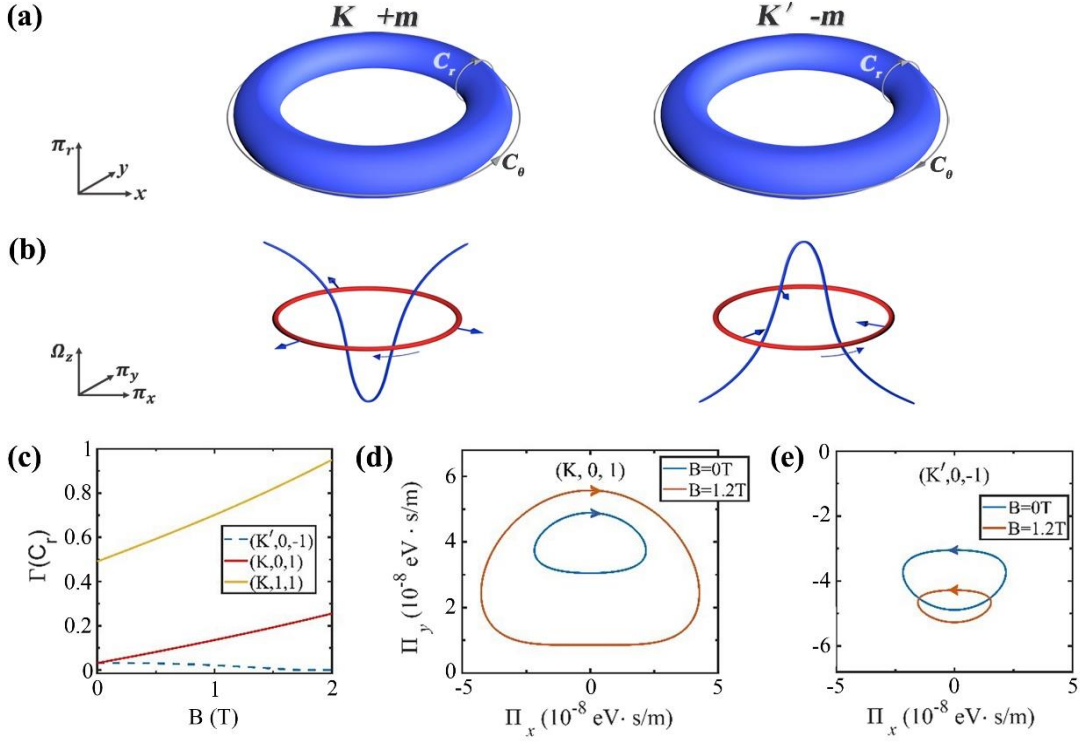

FIG. 4. (a) Closed loops $C_r$ and $C_\theta$ of the torus space for positive angular momentum state in the $K$ valley (left) and negative angular momentum state in the $K'$ valley (right). (b) Schematic of the momentum trajectories of the $C_r$ loop for the $(K, n, m)$ and $(K', n, -m)$ states (the red circles). The blue curves are the Berry curvature. Increasing the magnetic field, the momentum trajectories are pulled in the opposite directions for the $(K, n, m)$ and $(K', n, -m)$ states. (c) Berry phase as a function of the magnetic field for $(K, 0, 1)$, $(K', 0, -1)$ and $(K, 1, 1)$ states. (d, e) The momentum trajectories for $(K, 0, 1)$ and $(K', 0, -1)$ states at $B = 0$ T and 1.2 T. When $B = 0$ T, the momentum trajectories of the two valleys enclose the same area and the Berry phases acquired are the same. However, for nonzero $B$, the magnetic field can provide an effective Lorentz force that pulls the momentum trajectories outside (inside) for $m = \pm 1$ states, and increases (decreases) the Berry curvature flux, i.e. the Berry phase, as shown in panel (c).